\documentclass[aps,prd,10pt,amsmath,floats,floatfix,nofootinbib,superscriptaddress,preprintnumbers,twocolumn]{revtex4-2}

\usepackage{amsmath,amssymb,amsfonts,amsthm}
\usepackage{bm,bbm}
\usepackage{graphicx}
\usepackage[utf8]{inputenc}
\usepackage[table]{xcolor}
\usepackage[all]{xy}
\usepackage{pifont}
\usepackage{multirow}

\newcommand{\be}{\begin{equation}}
\newcommand{\ee}{\end{equation}}
\newcommand{\ba}{\begin{eqnarray}}
\newcommand{\ea}{\end{eqnarray}}
\def\bs{\begin{subequations}}
\def\es{\end{subequations}}
\newcommand{\bc}{\begin{center}}
\newcommand{\ec}{\end{center}}
\def\a{\alpha}
\def\b{\beta}

\def\g{\gamma}
\def\la{\lambda}
\def\k{\kappa}
\def\e{\epsilon}

\def\Om{\Omega}
\def\om{\omega}
\def\G{\Gamma}

\def\s{\sigma}

\def\N{\nabla}

\def\cD{\mathcal{D}}

\def\cF{\mathcal{F}}

\def\cL{\mathcal{L}}

\def\cN{\mathcal{N}}

\def\cR{\mathcal{R}}

\def\cV{\mathcal{V}}

\def\ds{d_{\rm S}}
\def\dh{d_{\rm H}}

\def\p{\partial}

\newcommand{\Eq}[1]{(\ref{#1})}

\def\cob{\color{blue}}

\newcommand{\au}[2]{#1.~#2}
\newcommand{\book}[5]{\emph{#1} (#2, #3, #4, #5)}
\newcommand{\books}[4]{\emph{#1} (#2, #3, #4)}
\newcommand{\oarX}[1]{\href{http://arxiv.org/abs/#1}{{\ttfamily\cob arXiv:#1}}}
\newcommand{\arX}[1]{\href{http://arxiv.org/abs/#1}{{\ttfamily\cob arXiv:#1}}}
\newcommand{\doin}[6]{\href{http://dx.doi.org/#1}{{\cob #2 #3 {\bf #4}, #5 (#6)}}}
\newcommand{\doinn}[5]{\href{http://dx.doi.org/#1}{{\cob #2 {\bf #3}, #4 (#5)}}}
\newcommand{\doij}[5]{\href{http://dx.doi.org/#1}{{\cob #2 #3 (#5) #4}}}

\newcommand{\proc}[5]{in \emph{#1}, edited by #2 (#3, #4, #5)}
\newcommand{\procsinm}{\proc}
\newcommand{\tia}[1]{#1,}

\def\tp{t_{\rm Pl}}
\def\rme{e}
\def\rmd{d}
\def\rmi{i}

\begin{document}

\title{Dark energy in multifractional spacetimes}

\author{Gianluca Calcagni}
\email{g.calcagni@csic.es}
\affiliation{Instituto de Estructura de la Materia, CSIC,\\ Serrano 121, Madrid, 28006 Spain}

\author{Antonio De Felice}
\email{antonio.defelice@yukawa.kyoto-u.ac.jp}
\affiliation{Center for Gravitational Physics, Yukawa Institute for Theoretical Physics,\\ Kyoto University, 606-8502, Kyoto, Japan}

\date{April 6, 2020}

\begin{abstract}
{We study the possibility to obtain cosmological late-time acceleration from a geometry changing with the scale, in particular, in the so-called multifractional theories with $q$-derivatives and with weighted derivatives. In the theory with $q$-derivatives, the luminosity distance is the same as in general relativity and, therefore, geometry cannot act as dark energy. In the theory with weighted derivatives, geometry alone is able to sustain a late-time acceleration phase without fine tuning, while being compatible with structure-formation and big-bang nucleosynthesis bounds. This suggests to extend the theory, in a natural way, from just small-scale to also large-scale modifications of gravity. Surprisingly, the Hausdorff dimension of spacetime is constrained to be close to the topological dimension 4. After arguing that this finding might not be a numerical coincidence, we conclude that present-day acceleration could be regarded as the effect of a new restoration law for spacetime geometry.}
\end{abstract}



\preprint{\doin{10.1103/PhysRevD.102.103529}{PHYSICAL REVIEW}{D}{102}{103529}{2020} [\arX{2004.02896}] \hspace{6cm} {\rm YITP-20-44}}

\maketitle


\section{Introduction}

To date, all cosmological observations agree on the fact that general relativity, incarnated in the so-called $\Lambda$CDM model, gives a valid and accurate description of the history of the universe and of the astrophysical objects within, from early times \cite{P18I,P18VI} until today \cite{Ab16b}. Despite the success reaped by Einstein theory, however, there is still room for theoretical speculation. In particular, late-time acceleration and the nature of dark energy are unsolved enigmas that do not quite fit in the picture. On one hand, even if cosmic acceleration is compatible with a plain cosmological constant $\Lambda$, its origin and small value do not find a convincing explanation in general relativity and ordinary particle physics, for several reasons (see, e.g., \cite[chapter 7]{CQC} for an overview on the cosmological constant problem). On the other hand, the value of the Hubble parameter today $H_0$ computed from the cosmic microwave background in the \textsc{Planck} Legacy release \cite{P18VI} is incompatible with the one found in local (Cepheids-based) observations \cite{Riess:2016jrr} by more than three standard deviations, a fact that might be imputed to the underlying assumption of the $\Lambda$CDM model.

Our aim in this work is to add another contender to the enormous amount of exotic dark-energy models already proposed in the literature, but with a new twist. It is well known that deviations from general relativity with a canonical scalar field can easily generate accelerating phases which, with the appropriate motivation, can provide an alternative explanation to dark energy. Performing this trick once again would only contribute to pile up phenomenological models while we wait for more precise measurements of the late-time expansion, for instance from \textsc{Euclid} \cite{Euc}. However, instead of proposing a phenomenological model of dark energy, here we take a class of gravitational theories, named multifractional spacetimes \cite{revmu,frc16}, originally presented as deviations from general relativity at the fundamental level and check what they have to say about late-time acceleration. This class of top-down approaches may be more appealing than \emph{ad hoc} models because the requirement they have to satisfy to recover a four-dimensional geometry at post-microscopic scales usually make their theoretical predictions at all scales rigid and, therefore, more easily falsifiable. This also happens for the macroscopic constraints on the models considered here, that will select one among several possibilities. The ensuing cosmological scenario will turn out to be worth investigating because it illuminates an unsuspected relation between late-time acceleration and spacetime geometry.

In all known theories of quantum gravity, the dimension of the effective spacetimes originated from quantum geometry changes with the probed scale, a phenomenon called dimensional flow related to the renormalizability of quantum gravity at short scales \cite{revmu,tH93,Car09,fra1,Car17}. Multifractional spacetimes implement this feature kinematically in the spacetime measure, which admits a universal parametrization \cite{revmu,first}, and in the dynamics via a model-dependent kinetic operator. Taken as independent theories, there are only four types of multifractional scenarios: with ordinary, weighted, $q$- or fractional derivatives in the dynamical action. The first three have received much attention \cite{revmu}, while the fourth, which we will ignore from now on, is still under construction \cite{revmu,Calcagni:2018dhp}. Unfortunately, the renormalizability of perturbative quantum gravity was shown \emph{not} to improve in the theories with $q$- and weighted derivatives \cite{revmu,frc9}, a result which indicated that these theories do not provide a fundamental microscopic description of Nature. However, the results of this paper will point out that one of these scenarios (the theory with weighted derivatives) can be a very interesting infrared (IR), rather than (or on top of) ultraviolet (UV), modification of gravity. It is important to stress that this extension of multifractional theories is admitted by a general theorem governing the form of the spacetime measure \cite{first} and, therefore, it is not introduced by hand. In fact, this holds for any theory of quantum gravity admitting dimensional flow in the Hausdorff dimension and a continuum approximation \cite{revmu,first}. The main implication of this result is that, even if the theories we will study here are not fundamental in the sense specified above, the feature responsible for late-time acceleration is expected to be present in other theories of quantum gravity with a scale-dependent Hausdorff dimension.\footnote{An example is the class of theories based on pre-geometric structures with discrete labels, such as group field theory and loop quantum gravity.} Therefore, our findings may be interesting not only because they complete the study of multifractional spacetimes and enrich their phenomenology to unsuspected directions, but also because they could be kept in mind when exploring the cosmology of quantum gravities with dimensional flow. The very idea to link dark energy to quantum gravity goes against the grain of effective field theory, since it is usually believed that quantum gravity can only have Planck-size corrections, which cannot possibly have an impact on the late-time universe. However, dimensional flow is essentially a nonperturbative phenomenon where other elements apart form the UV/IR divide enter the game, such as the discrete scale invariance responsible for the cosmic log-oscillations we will describe in due course.

The starting point for the study of the cosmology of these theories are the Friedmann--Lema\^itre--Robertson--Walker (FLRW) equations obtained from the full background-inde\-pend\-ent equations of motion \cite{fra2,frc11}. While there are some studies about inflation \cite{frc14,Calcagni:2017via} and gravitational-wave propagation \cite{qGW,Calcagni:2019ngc} that constrain or propose possible effects in the cosmic microwave background and in gravitational-wave signals in multifractional spacetimes, the problem of dark energy has received less attention, except in the case with ordinary derivatives. The absence of cosmological vacuum solutions in the theory with ordinary derivatives \cite{fra2} led some to explore scenarios with more or less exotic dark-energy components \cite{Karami:2012ra,Lemets:2012fp,Chattopadhyay:2013mwa,Maity:2016dfv,Jawad:2016piq,Sadri:2018lzz,Das:2018bxc,Debnath:2019isy} (see also \cite{Sheykhi:2013rla,Haldar:2016jdf} for a thermodynamical approach). However, our point of view is that a sensible theory beyond Einstein should be able to describe cosmic acceleration without introducing dark-energy fields by hand. In the case of multifractional spacetimes, this acceleration should come only from geometry \cite{frc11}. The theory with ordinary derivatives is less attractive than the others due to some technical issues \cite{revmu}; for this reason, we will turn to the theories with $q$-derivatives and with weighted derivatives, which admit a more rigorous structure.

In particular, we will show that geometry alone in the theory with $q$-derivatives is unable to sustain a dark-energy phase, contrary to the more interesting case with weighted derivatives. We will place constraints on the parameter space of this theory and find its compatibility with observations only when the Hausdorff dimension of spacetime is kept at the constant value $\dh=4$, i.e., the number of topological dimensions. We interpret this result as the manifestation of something we might call a restoration law, not apparent in previous works. It is as if late-time acceleration was the expression of the tendency of the Hausdorff spacetime dimension to stay close, or at least get back, to 4. The value of 4 is not conserved exactly but it is restored asymptotically. We give an argument why this might not be just a numerical coincidence, although it is too weak to establish a full-fledged physical principle yet.

In Secs.\ \ref{sec2a}--\ref{fra}, we briefly introduce some well-known cosmological quantities and review multifractional theories, quoting their main features on a homogeneous background. The reader interested in an in-depth discussion can consult \cite{revmu} and references therein. In Sec.\ \ref{exte}, we extend the spacetime measure of the theories to include beyond-general-relativistic correction terms, which will play a crucial role in the interpretation of our numerical results. The cosmological equations of the theory with $q$-derivatives are reviewed in Sec.\ \ref{frw}, while in Sec.\ \ref{noda} we show that geometry alone cannot accelerate the universe in this theory. This result is exact. The FLRW equations of the theory with weighted derivatives are presented in Sec.\ \ref{frw2}, while in Sec.\ \ref{fct} we encode all the effects of multiscale geometry into an effective dark-energy component obeying standard Friedmann equations. The numerical analysis and its discussion are presented in Secs.\ \ref{nume}--\ref{disc}, while Sec.\ \ref{conc} is devoted to conclusions.


\section{General setup}\label{sec2}


\subsection{Cosmology}\label{sec2a}

We work in four topological dimensions ($D=4$) and with signature $(-,+,+,+)$. The Hubble parameter is
\be
H := \frac{\dot{a}}{a}\,,
\ee
where a dot denotes a derivative with respect to cosmic time $t$. The evolution of the universe can be parametrized by the redshift $1+z=a_0/a$, where $a_0=a(0)=1$ is the scale factor today. We use a subscript 0 for any quantity evaluated today: the Hubble constant $H_0$, the age of the universe $t_0$, and so on.

The universe is assumed to be filled by perfect fluids with energy-momentum tensor
\be\label{pefl}
T_{\mu\nu} =(\rho+P)u_\mu u_\nu+g_{\mu\nu}\,P
\ee
and equation of state $P=w\rho$. In all models, the content of the universe will be nonrelativistic matter (dust, $\rho=\rho_{\rm m}$, $w=0$) plus a radiation component ($w=1/3$). Both curvature and the cosmological constant are set to zero, $\textsc{k}=0=\Lambda$, the first from observations \cite{P18I,P18VI} and the second because we want to get acceleration purely from geometry. The latter will give an effective contribution we will dub as $\rho_\textsc{de}$, where \textsc{de} stands for dark energy.

For each matter component, we will use the dimensionless energy-density parameter
\be\label{omm}
\Om_{\rm m}:=\frac{\rho_{\rm m}}{\rho_{{\rm crit}}}\,,\qquad \Om_\textsc{de}:=\frac{\rho_\textsc{de}}{\rho_{{\rm crit}}}\,,\qquad \rho_{{\rm crit}}:= \frac{3H^2}{\k^2}\,,
\ee
where $\k^2=8\pi G$ is Newton's constant. The late-time standard Friedmann equation in general relativity in the absence of curvature is
\be\label{hgr}
H\simeq H_0\sqrt{\Om_{{\rm m},0}\,(1+z)^3+\Om_{\textsc{de},0}\,(1+z)^{3(1+w_\textsc{de})}}\,,
\ee
for a dark-energy component with constant barotropic index $w_\textsc{de}$. This equation is modified in multifractional spacetimes, as we will see below.

To compare the theory with supernov\ae\ data, we will use the luminosity distance $d_L$, defined by imposing that the flux $\cF$ of light reaching an observer is the power $L$ emitted by a distant source per unit of area, measured on a sphere of radius $d_L$:
\be
\cF=:\frac{L}{4\pi d_L^2}\,.
\ee
This is a definition and it is completely model independent. The flux $\cF$ is measured from Earth, while $L$ is determined by the astrophysical properties of the source. Standard candles such as supernov\ae\ are sources for which $L$ is known. From this, one finds $d_L$. In turn, the luminosity distance can be expressed in terms of the dynamics and, in particular, of the dark-energy equation of state. It is this last part that depends on the specific cosmological model.

In standard general relativity, the luminosity distance a photon traveled from some source at redshift $z$ to Earth ($z=0$) is $a_0r$, as measured by the observed at $t_0$. Since $\rmd\tau^2=\rmd r^2$ on the light cone, $d_L$ is proportional to conformal time $\tau_0-\tau(z)$. Taking into account the redshift of power $L=\textrm{(energy)}/\textrm{(time)}\propto a/(1/a)=a^2$ of photons reaching the observer at different times, one gets \cite{CoLu}
\ba
d_L &=&\frac{a_0}{a}\,r = (1+z)[\tau_0-\tau(z)]\label{dl1}\\
    &=& (1+z)\int^{t_0}_{t(z)}\frac{\rmd t}{a}=(1+z)\int^{1}_{a(z)}\frac{\rmd a}{Ha^2}\nonumber\\
		&=&(1+z)\int_0^z\frac{\rmd z}{H}\,.\label{dl2}
\ea
Observations of standard candles such as type I supernov\ae\ allow one to determine the luminosity distance $d_L$, the local Hubble parameter $H_0$ and, from \Eq{hgr}, a constraint on the dark-energy barotropic index $w_\textsc{de}$.

In multifractional theories, the redshift law for time intervals and frequencies is the same and \Eq{dl1} does not change, but the relation between conformal time and the dynamics is modified and \Eq{dl2} receives several corrections, which we will discuss in due course. 


\subsection{Metric and geometric-harmonic structures}\label{sec2b}

Let $D=4$. Multifractional spacetimes are endowed with two mutually independent structures, a metric one and a geometric-harmonic one. The metric structure is given by the metric $g_{\mu\nu}$, in this work the flat FLRW metric with components
\be
g_{00}=-1\,,\qquad g_{ii}=a^2(t)\,,\quad i=1,2,3\,.
\ee

The geometric-harmonic structure is the integro-differential structure of the theory and is determined by the choice of spacetime measure
\be
\rmd^4 q(x)=\rmd^4x\,v(x)
\ee
in the action and by the type of derivatives in kinetic terms. In standard general relativity, $v=1$ and the action measure is $\rmd^4x$ (times the $\sqrt{|g|}$ volume weight, which is determined by the metric structure). The action measure weight $v(x)$ is the easiest to understand. The \emph{Hausdorff dimension} of spacetime is defined as the scaling of the Euclidean-signature volume $\cV(\ell)=\int_\textrm{ball,cube}^\ell\rmd^4x\,v(x)$ of a 4-ball or 4-cube with linear size $\ell$,
\be\label{dhdef}
\dh^{\rm spacetime}:=\frac{\rmd\ln\langle\cV(\ell)\rangle}{\rmd\ln\ell}\,,
\ee
where $\langle\cdot\rangle$ means average over the harmonic structure (see below). In ordinary spacetime, the volume scales as $\ell^4$, so that $\dh^{\rm spacetime}=4$. However, when the Hausdorff dimension of spacetime varies with the probed scale and this variation is slow and smooth at large scales or late times, its form is given by a unique parametrization \cite{revmu,first}, that we present here only in the time direction because we will be interested in homogeneous solutions on a homogeneous metric background. The full homogeneous measure weight
\be\label{vq}
v(t)=\dot q(t)
\ee
is, in the most general case, an infinite superposition of complex powers of time, $q(t)=\sum_l \g_l|t/t_l|^{\a_l+\rmi\om_l}$, where $\a_l,\om_l\in\mathbb{R}$ and $\g_l\in\mathbb{C}$ are dimensionless constants and $t_l\in\mathbb{R}$ are fundamental time scales of the geometry. However, reality of the measure constrains the parameters $\g_l$ and $\om_l$ of this sum to combine into a real-valued expression, a generalized polynomial deformed by logarithmic oscillations:
\bs\label{genpo}\ba
v(t) &=& \sum_{l=0}^{+\infty} \left|\frac{t}{t_l}\right|^{\a_{*,l}-1} F_l(t)\,,\\
F_l(t) &=& A_{0,l}+\sum_{n=1}^{+\infty} \tilde F_{n,l}(t)\,,\\
\tilde F_{n,l}(t) &=& A_{n,l}\cos\left(n\om_l\ln\left|\frac{t}{t_l}\right|\right)+B_{n,l}\sin\left(n\om_l\ln\left|\frac{t}{t_l}\right|\right),\nonumber\\\label{fnl}
\ea\es
where $l$ runs over the number of fundamental scales of geometry and $0<A_{n,l},B_{n,l}<1$ are constant amplitudes.


\subsection{UV binomial approximation}

The most common approximation of \Eq{genpo} used in physical applications includes only two terms, (i) one corresponding to the general-relativistic limit $v=1$ (obviously necessary to get viable phenomenology) and (ii) a correction term.
\begin{enumerate}
\item[(i)] Since, as we will see in Sec.\ \ref{fra}, the theory is defined in such a way that the dimensionality of spacetime coordinates is the one of our standard clocks and rulers ($[x^\mu]=-1$ in energy dimensions), then the $l=0$ term corresponding to the general-relativistic limit has $\a_{*,0}=1$. In other words, general relativity is the asymptotic IR limit of the theory.
\item[(ii)] In a scale hierarchy of spacetime made of only one scale $t_*$ ($l$ takes only two values $l=0,1$, where $t_0$ is not a physical scale because it is undetermined, since $\a_{*,0}=1$), the correction term is the UV limit of the measure. If other fundamental scales are also present ($l$ takes many values), then $t_*\equiv t_1$ corresponds to the shortest scale around which general relativity breaks down and UV anomalous scaling effects become visible.
\end{enumerate}
Taking only these terms and denoting $\a_*\equiv\a_{*,1}$, one gets the binomial (two-term generalized polynomial) expression
\be\label{bino}
v(t)\simeq 1+\left|\frac{t}{t_*}\right|^{\a_*-1}
\ee
in the absence of log oscillations, or 
\bs\label{logos}\ba
v(t) &\simeq& 1+\left|\frac{t}{t_*}\right|^{\a_*-1} F_\om(t)\,,\\
F_\om(t) &=& A_0+\sum_{n=1}^{+\infty} \tilde F_n(t)\,,\label{fn0}\\
\tilde F_n(t) &=& A_n\cos\left(n\om\ln\left|\frac{t}{t_*}\right|\right)+B_n\sin\left(n\om\ln\left|\frac{t}{t_*}\right|\right),\nonumber\\\label{fn}
\ea\es
in their presence. The positive parameter $\a_*$ is related to the Hausdorff dimension of spacetime in the UV,
\be\label{dh}
\dh^{\rm spacetime} \stackrel{\rm UV}{\simeq} \a_*+\sum_{i=1}^3\a_i\,,
\ee
where $\a_i$ are the fractional exponents in the spatial directions. The full measure is \rm$d^Dx\,v(x^0)\,\prod_{i=1}^{D-1}v_i(x^i)$, where each profile $v_i(x^i)$ along the spatial directions has the same parametric form as Eq.\ \Eq{logos} with $t$ replaced by $x^i$ and different parameters $\a_*\to \a_i$, $t_*\to\ell_i$, and so on. For a fully isotropic measure, $\a_*=\a_i$. In virtually all the literature on the subject, the range of values is chosen as $0<\a_*<1$ because these theories are thought of as UV modifications of ordinary field theories or of general relativity where, as in most quantum gravities, the dimension of spacetime decreases in the UV. 
 In this case, \Eq{dh} is the Hausdorff dimension of spacetime at early times $t\ll t_*$.

In the presence of log oscillations, the time scale inside the logarithms in Eq.\ \Eq{fn} was originally denoted as $t_\infty$ and regarded as independent from $t_*$. Usually, it was set to the Planck time $\tp\approx 5.3912 \times 10^{-44}\,\mbox{s}$ if $0<\a_*<1$, but later theoretical arguments led to the identification $t_\infty\equiv t_*$ \cite{first}. Therefore, in this paper we will not consider any such extra scale. The frequency
\be\label{omst}
\om=\frac{2\pi\a_*}{\ln N}\,,\qquad N=2,3,4,\dots\,,
\ee
takes one among a countable set of values, it appears as a consequence of a fundamental discrete scale invariance and parametrizes the typical time length of the discrete scale symmetry
\be
t\to \la_\om t:=\rme^{-\frac{2\pi}{\om}} t\,.
\ee
Thus, at scales of order of the log oscillations the time direction takes values on a deterministic fractal constructed with $N$ similarity maps with similarity ratios all equal to $\la_\om$ \cite{frc2}.

The constant $A_0$ in \Eq{fn0} is the zero mode of the modulation factor and can acquire two  values $A_0=0,1$ that can be understood as follows. The mesoscopic-to-large-scale form of the measure can be obtained with a coarse-graining procedure consisting in taking the average over a log oscillation \cite{NLM}. Restricting to a homogeneous background and denoting as 
\be\label{avef}
\langle f(t)\rangle:=\frac{1}{2\pi}\int_0^{2\pi}\rmd \nu\,f(\rme^{\frac{\nu}{\om}} t)
\ee
the average of a log oscillating function, one has
\be\label{mflr}
\langle F_\om\rangle=A_0\,,\qquad \langle F_\om^2\rangle=A_0^2+\sum_{n> 0}\frac{A_n^2+B_n^2}{2}\,.
\ee
The case $A_0=0$, never explored in the cosmological phenomenology of these theories, corresponds to a zero average: starting from mesoscopic temporal scales $t \gg \tp$, both the log-oscillatory pattern and the power-law contribution begin to disappear. In the case $A_0=1$, the only one considered in the cosmology literature so far \cite{frc11,frc14}, the power-law contribution survives longer, up to and beyond scales $t\gtrsim t_*$.


\subsection{Physical frame}\label{fra}

The form of the kinetic operators in the action is determined by the choice of symmetries and the requirement that, just like the Hausdorff dimension, in quantum gravity also the spectral dimension $\ds$ (the scaling of dispersion relations) changes with the probed scale. Thus, the system is defined by a nontrivial dimensional flow (varying $\dh$ and/or $\ds$) and certain action symmetries. In general, the metric is nonminimally coupled with matter fields due to the anomalous (i.e., nonconstant) geometric-harmonic structure, which leads to complicated equations of motion where Lorentz invariance is explicitly broken by the time scale $t_*$ (and the spatial scale $\ell_*$, which we do not consider here) appearing in the measure weight $v$. However, the problem is greatly simplified in two of the admissible cases, the theory with $q$-derivatives and the theory with weighted derivatives. We briefly recall their main points \cite{revmu}.

In the theory with $q$-derivatives, the original system is recast into a simpler one which is identical to the Standard Model of particle physics where all coordinates are the geometric profiles $q^\mu(x^\mu)$. When gravity is turned on, the original system becomes equivalent to general relativity in $q$ coordinates, plus matter fields. To perform calculations, one can temporarily forget that the $q^\mu$ are composite coordinates and work in this \emph{integer frame} until the construction of physical observables, at which point one must revert to the physical (or \emph{fractional}) frame where the geometry is multiscale. In this theory, the fractional frame is such that clocks and rods do not adapt with the observation scale, while they do in the integer frame. Consequently, in the fractional frame we can observe dimensional flow through its imprint on certain physical observables, while in the artificial integer frame the geometry is constant. The mapping $x^\mu\to q^\mu(x^\mu)$ connects the two frames and is not a coordinate transformation in the sense of general relativity.

In the theory with weighted derivatives, the integer frame is defined by field transformations $\phi_{\mu\nu\cdots}\to\tilde\phi_{\mu\nu\cdots}(\phi,v)$ involving $v$. In the absence of gravity, the electroweak-strong model of quantum interactions in the fractional frame is mapped into the standard Standard Model on Minkowski spacetime. In the presence of gravity, the dynamics in the fractional frame is very similar to a scalar-tensor theory where $v$ plays the role of the scalar, but only in a superficial visual analogy at the level of the equations of motion: $v$ is not a field, since it is given by the nondynamical fixed profile \Eq{genpo} and its spatial generalization. In the unphysical integer frame, the theory is mapped into general relativity plus a purely geometric, $v$-dependent contribution.

We insist that in these theories the integer and fractional frame are not physically equivalent and that we live in the latter, while the integer frame is just a convenient trick to simplify calculations. The multifractional theory with fractional derivatives does not admit a frame mapping, which is the reason why it has been studied much less than the other two. In general, having a multiscale geometry selects a preferred frame, at least in multifractional theories. Other multiscale scenarios such as string theory, asymptotically safe quantum gravity, nonlocal quantum gravity and group field theory (which includes spin foams and loop quantum gravity) have a higher degree of symmetry (Lorentz and diffeomorphism invariance are preserved or deformed in a controlled way) and do not require a frame choice, once a low-energy or semi-classical notion of spacetime is recovered.


\subsection{Infrared extension of the measure}\label{exte}

Equations \Eq{bino} and \Eq{logos} are the profiles typically considered in the literature. The constant unit term corresponds to the measure at the scales on which general relativity holds, hence its normalization to 1. The physical picture usually advocated is that multiscale spacetimes and the dynamics therein are anomalous at short scales $t\lesssim t_*$ and reduce to ordinary spacetimes and general relativity at large scales $t\gg t_*$. Therefore, the general-relativistic regime and its measure weight $v=1$ constitute the terminal regime of dimensional flow at large scales.

However, this is only one of the possible applications of the dimensional-flow theorem \cite{first} giving rise to the most general profile \Eq{genpo}. In principle, ultra-IR terms dominating the measures at scales larger than those of the general-relativistic regime are possible. For example, ignoring log oscillations for the sake of the argument, the measure weight \Eq{bino} can be replaced by a three-term profile
\be\label{bino2}
v(t)\simeq \left|\frac{t}{t_*}\right|^{\a_*-1}+1+\left|\frac{t}{t_{\rm c}}\right|^{\a_{\rm c}-1},
\ee
where $t_*$ is close to Planck time $\tp$ and $0<\a_*<1$, while $t_{\rm c}\gg t_*$ and
\be\label{priora}
\a_{\rm c}>1\,.
\ee
The subscript ``c'' stands for cosmological. According to this extended measure, general relativity is only a transient regime between a deep-UV limit, where microscopic quantum-gravity effects are important, and an ultra-IR regime which deviates from standard Einstein theory at cosmological scales. In other words, our clocks (and rulers, in an inhomogeneous setting) not only changed in the past, but they will also change in the future. We will see that dark energy can be interpreted as a manifestation of this deviation.

Therefore, here we will explore a model where the correction to general relativity is not a UV term but a ``post-IR'' one. The UV correction, with $0<\a_*<1$ and $t_*\sim \tp$, is totally negligible at late times and can be ignored. The final expression for the measure is therefore identical to \Eq{logos} but with a new set of parameters with prior \Eq{priora}:
\bs\label{logos2}\ba
v(t) &\simeq& 1+\left|\frac{t}{t_{\rm c}}\right|^{\a_{\rm c}-1} F_\om(t)\,,\\
F_\om(t) &=& A_0+\sum_{n=1}^{+\infty} \tilde F_n(t)\,,\\
\tilde F_n(t) &\simeq& A_n\cos\left(n\om\ln\left|\frac{t}{t_{\rm c}}\right|\right)+B_n\sin\left(n\om\ln\left|\frac{t}{t_{\rm c}}\right|\right),\nonumber\\\label{fn2}\\
\om&=&\frac{2\pi\a_{\rm c}}{\ln N}\,,\qquad N=2,3,4,\dots\,.\label{omspe}
\ea\es

In this paper, we will consider the following cases in increasing order of difficulty:
\begin{enumerate}
\item[\textbf{1}.] Binomial measure without log oscillations:
\be\label{nolog}
A_0=1\,,\qquad \tilde F_n(t)=0\,.
\ee
\item[\textbf{2}.] Binomial measure with log oscillations up to a finite number of harmonics:
\be
F_\om(t) = A_0+\sum_{n=1}^{n_{\rm max}}\tilde F_n(t)\,.
\ee
\begin{enumerate}
\item[\textbf{2a}.] One harmonic ($n_{\rm max}=1$) with $N=2$ [smallest $N$ in \Eq{omspe}] and $A_0$, $A_1$ and $B_1$ free parameters. In turn, this and the following can be divided into two subcases, $A_0=0$ (no zero mode) and $A_0=1$.
\item[\textbf{2b}.] One harmonic with very large $N$. We choose the arbitrary value $N=\rme^{10}$ to get a suppression factor $1/10$ in $\om$ (small frequency).
\end{enumerate}
\item[\textbf{3}.] Binomial measure with many harmonics ($n_{\rm max}>1$).
\begin{enumerate}
\item[\textbf{3a}.] Many harmonics with $N=2$. Different theoretical choices for the $n$-dependence of $A_n$ and $B_n$ indicate that higher modes are rapidly suppressed and that it is sufficient to consider $n_{\rm max}\lesssim 10$ \cite{Calcagni:2017via}. We take $n_{\rm max}= 10$. Also, to maximize the effect we take all amplitudes constant and equal to one another: $A_1=A_2=\ldots=A_{n_{\rm max}}$ and $B_1=B_2=\ldots=B_{n_{\rm max}}$. This configuration can be regarded as a multiharmonic measure where only the first $O(10)$ harmonics dominate.
\item[\textbf{3b}.] Many harmonics with $N=\rme^{10}$.
\end{enumerate}
\end{enumerate}


\section{Theory with $q$-derivatives}\label{qder}


\subsection{Action and FLRW dynamics}\label{frw}

In the theory with $q$-derivatives, the dynamical action is the same as in general relativity, except that all the coordinates $x^\mu$ are replaced by the multifractional profile $q^\mu(x^\mu)=\int \rmd x^\mu\,v_\mu(x^\mu)$, where the index $\mu$ is not contracted. As we commented in Sec.\ \ref{fra}, this is not a coordinate transformation because measurement units differ in the fractional and integer frame. Physical clocks and rulers, used for actual measurements, are defined on the manifold spanned by the coordinates $x^\mu$. In this fractional frame, the action measure $\rmd^D q(x)=\prod_{\mu=0}^{D-1}\rmd q^\mu(x^\mu)$ reflects the basic postulate of the theory, namely, that spacetime geometry changes with the scale. On the other hand, in the integer frame the $q^\mu$ are regarded as noncomposite coordinates and, therefore, the spacetime measure is the standard Lebesgue measure $\rmd^D q$, with no dimensional flow.

In the fractional frame, the gravitational action in the absence of a cosmological constant is \cite{frc11}
\be\label{Sgq}
S_q =\frac{1}{2\k^2}\int\rmd^Dx\,v\,\sqrt{|g|}\,\,{}^q R+S_{\rm matter}\,,
\ee
where the Ricci scalar ${}^q R:= g^{\mu\nu}\, {}^q R_{\mu\nu}$, the Ricci tensor ${}^q R_{\mu\nu}:= {}^q R^\rho_{~\mu\rho\nu}$ and the Riemann tensor
\ba
{}^q R^\rho_{~\mu\s\nu} &:=& \frac{1}{v_\s}\p_\s {}^q\G^\rho_{\mu\nu}-\frac{1}{v_\nu}\p_\nu {}^q\G^\rho_{\mu\s}\nonumber\\
&&+{}^q\G^\tau_{\mu\nu}\,{}^q\G^\rho_{\s\tau}-{}^q\G^\tau_{\mu\s}\,{}^q\G^\rho_{\nu\tau}\,,\nonumber\\
{}^q\G^\rho_{\mu\nu} &:=& \tfrac12 g^{\rho\s}\left(\frac{1}{v_\mu}\p_{\mu} g_{\nu\s}+\frac{1}{v_\nu}\p_{\nu} g_{\mu\s}-\frac{1}{v_\s}\p_\s g_{\mu\nu}\right)\,,\nonumber
\ea
are all made with $q$-derivatives (no summation over $\mu$)
\be
\frac{\p}{\p q^\mu(x^\mu)}=\frac{1}{v_\mu(x^\mu)}\frac{\p}{\p x^\mu}\,.
\ee
The matter action is constructed according to the same criterion.

The Friedmann equations in $D=4$ topological dimensions in the presence of a perfect fluid are \cite{frc11}
\ba
H^2&=&\frac{\k^2}{3}\,v^2\rho\,,\label{fri}\\
\frac{\ddot a}{a}&=&-\frac{\k^2}{6}\,v^2(\rho+3P)+H\frac{\dot v}{v}\,.\label{ra}
\ea
The continuity equation for a perfect fluid is compatible with the Friedmann equations:
\ba\label{ceq}
\dot\rho+3H(\rho+P)=0\,.
\ea
Multiscale geometry effects may be implicit in the energy density and pressure. For instance, for a homogeneous scalar field
\be
\rho_\phi=\frac1{2v^2}\dot\phi^2+W(\phi)\,,\qquad P_\phi=\frac1{2v^2}\dot\phi^2-W(\phi)\,,
\ee
and its continuity equation is modified both in the friction and in the potential term:
\be
\ddot\phi+\left(3H-\frac{\dot v}{v}\right)\dot\phi+v^2 W_{,\phi}=0\,.\label{frira3}
\ee


\subsection{No dark energy from geometry}\label{noda}

In the multifractional theory with $q$-derivatives, the geometry of spacetime is such that it eases the slow-roll condition in inflation \cite{frc11,frc14} and, similarly, it could enhance the (usually insufficient) acceleration triggered by a quintessence field, i.e., it could relax the fine tuning in the initial conditions. However, we do not wish to straighten a general-relativistic model with some extra exotic ingredient. A genuine alternative explanation of dark energy should work without relying on a quintessential component. Therefore, we would like to get late-time acceleration from pure geometry. 

Unfortunately, this is not possible in the theory with $q$-derivatives because geometry effects cancel out in the luminosity distance. The latter receives two corrections, one from the definition of conformal time and one from the dynamics. Distances, areas and volumes all must be calculated with the nontrivial multiscale measure weight $v(x)$ and, in particular, integration along the time direction includes a factor $v(t)$. In particular, the luminosity distance $d_L$ is still proportional to the comoving distance (times the same redshift factor as in general relativity), but now the latter is $\int^0_{-q(r)} \rmd q(\bm{x})$. The FLRW line element
\ba
\rmd s^2&=&g_{\mu\nu}\,\rmd q^\mu(x^\mu)\,\rmd q^\nu(x^\nu)=-\rmd q^2(t)+a^2\,\rmd q^2(\bm{x})\nonumber\\
&=&-v^2(t)\,\rmd t^2+a^2v^2(\bm{x})\,\rmd \bm{x}^2\,,
\ea
where $v(\bm{x}):=v_1(x^1)v_2(x^2)v_3(x^3)$ is the spatial measure factorized in the coordinates, vanishes for light rays, so that the comoving distance is given by the integration of the weighted conformal time $\rmd t\,v(t)/a$.\footnote{Note that this time redefinition does not make the line element conformally flat, unless $v(\bm{x})=1$. However, we keep the misnomer ``conformal'' for the sake of an easier comparison with general relativity.} Thus, the luminosity distance is
\be
d_L=(1+z)\int^{t_0}_{t(z)}\frac{\rmd t\,v(t)}{a}=(1+z)\int_0^z\frac{\rmd z\,v}{H}\,.
\ee
On the other hand, the Friedmann equation with only dust and no dark-energy component reads
\be
H=v\,H_0\sqrt{\Omega_{{\rm m},0}\,(1+z)^3}\,,
\ee
so that
\be
d_L=\frac{1+z}{H_0}\int_0^z\frac{\rmd z}{\sqrt{\Omega_{{\rm m},0}\,(1+z)^3}}\,,
\ee
exactly as in the cold dark matter (CDM) model of Einstein gravity without cosmological constant. The conclusion is that this expression cannot possibly fit supernov\ae\ data and that the theory with $q$-derivatives cannot make the late-time universe accelerate just from pure geometry.


\section{Theory with weighted derivatives}\label{wder}

For the cosmologist, the theory with weighted derivatives may result more intuitive than the previous one because here the line element is the usual of general relativity and the only change in the expression for the luminosity distance \Eq{dl2} is in the profile $H(z)$.


\subsection{Action and FLRW dynamics}\label{frw2}

Weighted derivatives are ordinary derivatives with measure-weight factors inserted to the left and to the right:
\be\label{bD}
\cD:=\frac{1}{v^\b}\p(v^\b\,\cdot\,)\,,\qquad \b=\frac{2}{D-2}\,.
\ee
The gravitational action in the fractional (physical) frame is \cite{frc11}\footnote{Here $U$ has been rescaled by a factor of $2(D-1)$ with respect to \cite{frc11,revmu} in order to make the Friedmann equations simpler.}
\ba
S_v &=&\frac{1}{2\k^2}\int\rmd^Dx\,v\,\sqrt{|g|}[\cR-\g(v)\cD_\mu v\cD^\mu v\nonumber\\
&&\qquad-2(D-1)\,U(v)]+S_{\rm matter}\,,\label{eha}
\ea
where $\g$ and $U$ are functions of the weight $v$, $S_{\rm matter}$ is the minimally coupled matter action, $\cR:= g^{\mu\nu}\cR_{\mu\nu}$, $\cR_{\mu\nu}:= \cR^\rho_{~\mu\rho\nu}$ and
\ba
\cR^\rho_{~\mu\s\nu} &:=& \p_\s {}^\b\G^\rho_{\mu\nu}-\p_\nu {}^\b\G^\rho_{\mu\s}+{}^\b\G^\tau_{\mu\nu}{}^\b\G^\rho_{\s\tau}-{}^\b\G^\tau_{\mu\s}{}^\b\G^\rho_{\nu\tau}\,,\nonumber\\
{}^\b\G^\rho_{\mu\nu} &:=& \tfrac12 g^{\rho\s}\left(\cD_{\mu} g_{\nu\s}+\cD_{\nu} g_{\mu\s}-\cD_\s g_{\mu\nu}\right)\,.\nonumber
\ea
To avoid confusion with ordinary curvature tensors $R_{\dots}$, we used the curly symbol $\cR_{\dots}$ to denote tensors written in terms of the derivative \Eq{bD}. The metric is covariantly conserved with respect to the weighted covariant derivative $\N_\s^- g_{\mu\nu} := \p_\s g_{\mu\nu}-{}^\b\G_{\s\mu}^\tau g_{\tau\nu}-{}^\b\G_{\s\nu}^\tau g_{\mu\tau}=0$, implying that these spacetimes are Weyl integrable and that the metric is not conserved in the ordinary sense, $\N_\s g_{\mu\nu}=(\b\p_\s\ln v)\,g_{\mu\nu}$.

As we said in Sec.\ \ref{fra}, in the integer frame the theory resembles a scalar-tensor model in the Einstein frame. Denoting with a bar coordinates and metric quantities evaluated in this frame, the Friedmann equations in $D=4$ dimensions are \cite{frc11}
\ba
\bar H^2&=&\frac{\k^2}{3}\,\bar \rho+\frac{\Om}{2}\frac{(\p_{\bar t} v)^2}{v^2}+\frac{U(v)}{v}\,,\label{friw}\\
\frac{\p_{\bar t}^2\bar a}{\bar a}&=&-\frac{\k^2}{6}\,(\bar\rho+3\bar P)+\frac{U(v)}{v}\,,\label{friw2}
\ea
where derivatives are with respect to time $\bar t$ and $\Om=-3/2+9v^2\g(v)/4$ is a function of the measure weight. One can get the second Friedmann equation \Eq{friw2} either directly from the equations of motion or by using the continuity equation
\ba\label{ra2}
\p_{\bar t}\bar\rho+3\bar H(\bar\rho+\bar P)=0\,.
\ea
While, in general, $U$ is required to be nonzero for consistency of cosmological solutions, the function $\g$ was originally introduced to make an analogy with scalar-tensor models, with the difference that $v$ is not a dynamical field. However, this term is neither dictated by symmetries nor by consistency of the solutions and it can be dropped without loss of generality or creating any theoretical or conceptual problem. Since our goal is to test the most minimalistic model with weighted derivatives, i.e., with log-oscillating measure and no kinetic-like term, we can set $\g$ to vanish, so that $\Om=-3/2$ (this was done only in the numerical code; the equations written here are valid in general).

Two major points of departure with respect to scalar-tensor theories are, first, that $v$ is not a scalar field but a time profile fixed \emph{a priori} [by \Eq{genpo} in the fractional frame; see below]. And, second, that the ``potential'' term $U(v)$ is neither chosen \emph{ad hoc} nor reconstructed from observations, but it is determined from the dynamics itself. In fact, combining \Eq{friw} and \Eq{friw2} to eliminate the last term in the right-hand side, one gets the master equation
\be\label{master}
\frac{\p_{\bar t}^2\bar{a}}{\bar a}-\bar H^2+\frac{\k^2}{2}\,(\bar\rho+\bar P) =-\frac{\Om}{2}\frac{(\p_{\bar t} v)^2}{v^2}\,.
\ee
Thus, given the matter content one obtains $\bar a(\bar t)$ and, from \Eq{friw} or \Eq{friw2}, $U(v)$.

These equations are in the integer (Einstein) frame, which is unphysical in this theory. The dynamical equations in the physical (fractional, Jordan) frame, which we write here for the first time, are obtained by recalling that the Jordan and Einstein metric are related to each other by
\be\label{cori}
\bar g_{\mu\nu}=v\,g_{\mu\nu}\,,\qquad \bar a=\sqrt{v}\,a\,,
\ee
where $v=v(t)$ is the temporal part of the measure weight. Also, having chosen the FLRW metric in the integer frame implies that the same metric holds in the fractional frame provided $-\rmd \bar t^2=-v\rmd t^2$, i.e., 
\be
\frac{\rmd}{\rmd \bar t}=\frac{1}{\sqrt{v}}\frac{\rmd}{\rmd t}\,,
\ee
hence
\ba
&&\bar H = \frac{1}{\sqrt{v}}\left(H+\frac{1}{2}\frac{\dot v}{v}\right)\,,\nonumber\\
&&\frac{1}{\bar a}\frac{\rmd^2 \bar{a}}{\rmd\bar t^2}=
\frac{1}{v}\left[\frac{\ddot a}{a}+\frac12\left(\frac{\ddot v}{v}+H\frac{\dot v}{v}-\frac{\dot v^2}{v^2}\right)\right],
\ea
where dots are derivatives with respect to $t$. To these expressions, one must add those for the energy density and pressure of the perfect fluid:
\be
\bar\rho=\frac{\rho}{v^2}\,,\qquad \bar P=\frac{P}{v^2}\,,
\ee
where the factor $1/v^2$ stems from the conformal rescaling \Eq{cori}, $\overline{\textrm{(energy)}}/\overline{\textrm{(volume)}}=[\textrm{(energy)}$ $/\sqrt{v}]/[v^{3/2}\textrm{(volume)}]$.

Therefore, the Friedmann equations and the master equation in the physical frame read
\ba
\hspace{-.2cm}&&H^2+H\frac{\dot v}{v}=\frac{\k^2}{3v}\,\rho+\frac12\left(\Om-\frac{1}{2}\right)\frac{\dot v^2}{v^2}+U,\label{friw2w}\\
\hspace{-.2cm}&&\frac{\ddot a}{a}+\frac12\left(\frac{\ddot v}{v}+H\frac{\dot v}{v}-\frac{\dot v^2}{v^2}\right)=-\frac{\k^2}{6v}\,(\rho+3P)+U,\label{friw22}\\
\hspace{-.2cm}&&\dot H=-\frac{\k^2}{2v}\,(\rho+P)+\frac12\left[\left(\frac{3}{2}-\Om\right)\frac{\dot v^2}{v^2}-\frac{\ddot v}{v}+H\frac{\dot v}{v}\right].\label{master2}
\ea
It is in these equations, not in \Eq{friw}--\Eq{master}, where $v(t)$ is given by \Eq{vq} and \Eq{logos2}. Notice that the contributions of the measure weight mimic both a running effective background-evaluated Newton's constant
\be
G_{\rm eff}= \frac{G}{v}<G
\ee
and a phantom ``scalar'' with negative kinetic energy. Combined together, these features can sustain cosmic acceleration, without the theoretical inconvenience of phantom fields ($v$ is not a dynamical degree of freedom and is not associated with classical instabilities or negative-norm quantum states).

Finally, the continuity equation \Eq{ra2} becomes
\be\label{ceq2}
0 =\dot\rho+3H(\rho+P)-\frac{1}{2}\frac{\dot v}{v}(\rho-3P)\,.
\ee
Radiation ($w=1/3$), which is the only conformal invariant perfect fluid, is conserved also in this frame, while any other fluid experiences an effective dissipation.


\subsection{Effective dark-energy component}\label{fct}

The running of the Hausdorff dimension $\dh$ in this theory can drive one or more phases of acceleration. It is convenient to recast the contribution of the multiscale geometry and dynamics as a dark-energy component in standard general relativity. Thus, the first Friedmann equation \Eq{friw2w} is written as
\bs\label{frgr0}\ba
H^2&=&\frac{\k^2}{3}(\rho_{\rm m}+\rho_\textsc{de})\,,\\
\rho_\textsc{de}&:=&\frac{3}{\k^2}\left[\frac12\left(\Om-\frac{1}{2}\right)\frac{\dot v^2}{v^2}+U-H\frac{\dot v}{v}\right]-\left(1-\frac{1}{v}\right)\rho_{\rm m}\,,\nonumber\\
\ea\es
or, equivalently, as
\be\label{frgr}
\Om_{\rm m}=1-\Om_\textsc{de}\,,
\ee
where the dimensionless energy densities were defined in \Eq{omm}. This equation replaces \Eq{hgr}. Since dust matter does not obey the standard continuity equation, $\Om_{\rm m}\neq \Om_{{\rm m},0}(1+z)^3$. The effective dark-energy pressure $P_\textsc{de}$ is obtained by reformulating the master equation \Eq{master2} as its Einstein-gravity counterpart:
\ba
\hspace{-.2cm}\dot H&=&-\frac{\k^2}{2}(\rho_{\rm m}+\rho_\textsc{de}+P_\textsc{de})\,,\nonumber\\
\hspace{-.2cm}P_\textsc{de}&:=&-\frac{1}{\k^2}\left[\frac12\left(\frac32+\Om\right)\frac{\dot v^2}{v^2}-2H\frac{\dot v}{v}-\frac{\ddot v}{v}+3U\right],\label{pres}
\ea
which allows us to find the dark-energy barotropic index
\be\label{wDE}
w_\textsc{de}:=\frac{P_\textsc{de}}{\rho_\textsc{de}}\,,
\ee
a complicated function of the Hubble parameter, the measure weight $v$ and its time derivatives. For $\Om=-3/2$, the first term in the pressure \Eq{pres} vanishes. Note that \Eq{wDE} differs from the effective barotropic index coming from the contribution of all (matter and dark energy) components, defined as $w_{\rm eff}:=-1+2\e/3$, where
\ba
\hspace{-.4cm}\e&:=&-\frac{\dot H}{H^2}\nonumber\\
\hspace{-.4cm}&=&\frac{3}{2v}(1-\Om_\textsc{de})+\frac{\ddot v-H\dot v}{2H^2v}-\frac12\left(\frac32-\Om\right)\frac{\dot v}{H^2v^2}\label{eps}
\ea
is the first slow-roll parameter. Another quantity of interest is the deceleration parameter {\rm q} (not to be confused with the composite coordinates $q$)
\be
{\rm q}:= -\frac{\ddot{a}}{aH^2}=-1+\e \,,
\ee
related to $w_{\rm eff}$ by $w_{\rm eff}=(-1+2{\rm q})/3$.


\subsection{Solving the dynamics}\label{nume}

The dynamics can be solved numerically from a minimal set of differential equations. We choose the number of e-foldings as the main time variable, together with other dimensionless varying or constant parameters:
\ba
&&\cN:=\ln \frac{a_0}{a}\,,\qquad y:=\frac{t}{t_0},\qquad \mathfrak{h}:=\frac{H}{H_0}\,,\nonumber\\
&&b:=t_0H_0\,,\qquad y_{\rm c}:=\frac{t_{\rm c}}{t_0}=\frac{t_{\rm c}\,H_0}{b}.
\ea
Since $v$ is a given function of time, it is also a given function of $y$. Therefore, one could consider $y$ as the independent variable of integration. However, we find it useful to have an autonomous system of differential equations. In this case, we consider $\mathcal{N}$ to be the independent variable. On doing so, the system of equations can be written as
\ba
\mathfrak{h}' &=& -\frac{2\,b\mathfrak{h}vv_{,y}-6\,b^{2}\mathfrak{h}^{2}v\Omega_{\rm m}-2\,v_{,y}^{2}\Omega-2\,vv_{,{\it yy}}+3\,v_{,y}^{2}}{4b^{2}v^{2}\mathfrak{h}}\,,\nonumber\\\label{eq1}\\
\Omega_{\rm m}' &=& \frac{\Omega_{\rm m}}{2b^{2}\mathfrak{h}^{2}v^{2}}\left(6\,b^{2}\mathfrak{h}^{2}v^{2}-6\,b^{2}\mathfrak{h}^{2}v\Omega_{\rm m}+b\mathfrak{h}vv_{,y}\right.\nonumber\\
&&\qquad\qquad\left.-2\,v_{,y}^{2}\Omega-2\,vv_{,{\it yy}}+3\,v_{,y}^{2}\right)\,,\\
y' &=&-\frac{1}{b\mathfrak{h}}\,,\label{eq3}
\ea
where a prime denotes the derivative with respect to $\cN$. We also remind the reader that here we will consider $\Omega=-3/2$. In the most general case \Eq{genpo}, several dimensionless
parameters appear, while in the trinomial case \Eq{bino2} $v(y)=1+b_{\rm c} y^{\a_{\rm c}-1}+\cdots$, where the ellipsis is the UV correction, negligible at cosmic scales. This is the profile $v(y)$ considered here, $y$ being a function of $\cN$. Among the three dynamical equations \Eq{eq1}--\Eq{eq3}, the first corresponds to the second (modified) Einstein equation, whereas the
second one is determined by using the continuity equation. Finally, the third equation is a direct consequence of the definition of the e-folds number. It is also simple to show that
\ba
{\rm q} &=& -1+\frac{\mathfrak{h}'}{\mathfrak{h}}\,,\\
w_\textsc{de}&=&\frac{1-2{\rm q}}{3\,(\Omega_{\rm m}-1)}\,,
\ea
We can integrate the equations of motion from $\cN=0$ up to $\cN=6$, giving the following initial conditions:
\be
y(0) =1\,,\qquad \mathfrak{h}(0) =1\,,\qquad \Omega_{\rm m}(0) =\Omega_{{\rm m}0}\,.
\ee

In any case, we find that the system goes to matter domination at $\cN=6$, i.e., $\Omega\to1$ and ${\rm q}\to1/2$. Further equations are for $z$ and for the luminosity distance:
\be
\tilde d'=\tilde d+\frac{(1+z)^2}{\mathfrak{h}}\,,\qquad \tilde d:=H_0d_L\,,\qquad z' = 1+z\,.
\ee

Let us give some time scales for reference. $t_0\approx 14\times 10^9\,{\rm yr}\approx 10^{17}\,{\rm s}$ is the age of the universe today and $H_0=H(t_0)$. The onset of big-bang nucleosynthesis is at $t\approx 200\,{\rm s}$, corresponding to $y\approx 4.5\times 10^{-17}$ and $\cN\approx 21$, while matter-radiation equality happens at $t\approx 7\times 10^4\,{\rm yr}$, $y\approx 5\times 10^{-6}$, $\cN\approx 8$. Our numerical integration starts from today and ends at $\cN=6$, at the peak of matter domination. This choice is due to having focused our attention only to late-time data and, in particular, to data which do not need a description in terms of perturbation dynamics. In particular, we will use the constraints from type Ia supernov\ae\ (Union 2.1 set) \cite{Suzuki:2011hu}, today's value of $H_0$ (given in \cite{Riess:2016jrr}) and, finally, the constraint on the age of the universe $t_0$ coming from studying the globular cluster NGC 6752 \cite{Renzini}. This choice of data sets is given by the fact that we only focus on the late-time evolution of the models we are considering. High-redshift information, both from the background (necessary, e.g., for baryon acoustic oscillations) and perturbation sides (necessary, e.g., for \textsc{Planck} data), require an investigation of the multifractional theory which goes beyond the scope of this study. In particular, we want to see if the model introduced here could be compatible with data which only require knowledge of the dynamics at low redshift. As we will show, although the above data sets are limited, they will be enough to constrain considerably the parameter space of our model. The reason is that the acceleration of the universe is realized by means of the measure $v(y)$, but the latter does not represent itself a dark-energy component, i.e., a cosmological constant. Since the function $v(y)$ is given \emph{a priori} by the dimensional-flow theorem \cite{first}, it is not surprising that some of its approximated forms may not succeed to make the universe accelerate.

Regarding the time scales appearing in $v(y)$, the UV scale $t_*$ received strong bounds from several observations and experiments \cite{revmu,frc16}, ranging from Standard Model forces time scales to the Planck scale:
\be
10^{-61}\leq y_*=\frac{t_*}{t_0}\leq 10^{-47}\,. \label{tstt0}
\ee
The lower bound in \Eq{tstt0} corresponds to the Planck scale $t_*=\tp$, while the upper bound was obtained under the assumption that $0<\a_*<1$ \cite{frc16}. 
 These constraints apply only to the UV part of the trinomial measure weight \Eq{bino2}, in a regime (particle-physics scales) where the ultra-IR correction in \Eq{bino2}, the most important correction in the cosmological model we will consider here, is negligible. Conversely, the UV correction at near-Planckian scales $\sim t_*$ is completely negligible at late times and we will ignore it here.

The general procedure we followed to obtain the numerical constraints below was to sample the background-evolution parameter space for the chosen data sets. We performed an MCMC sampling on the parameters of the theory, letting the sampler (EMCEE \cite{emcee}) finding the minima of the $\chi^2$ distribution together with the 2$\sigma$ constraints for each parameter. In order to double check the MCMC results, we also performed a numerical minimization of the $\chi^2$ (via different methods such as the Newton one) and verified that results were compatible with the MCMC sampling.


\subsection{Results}\label{resu}

\begin{figure*}[t]
\bc
\includegraphics[width=15cm]{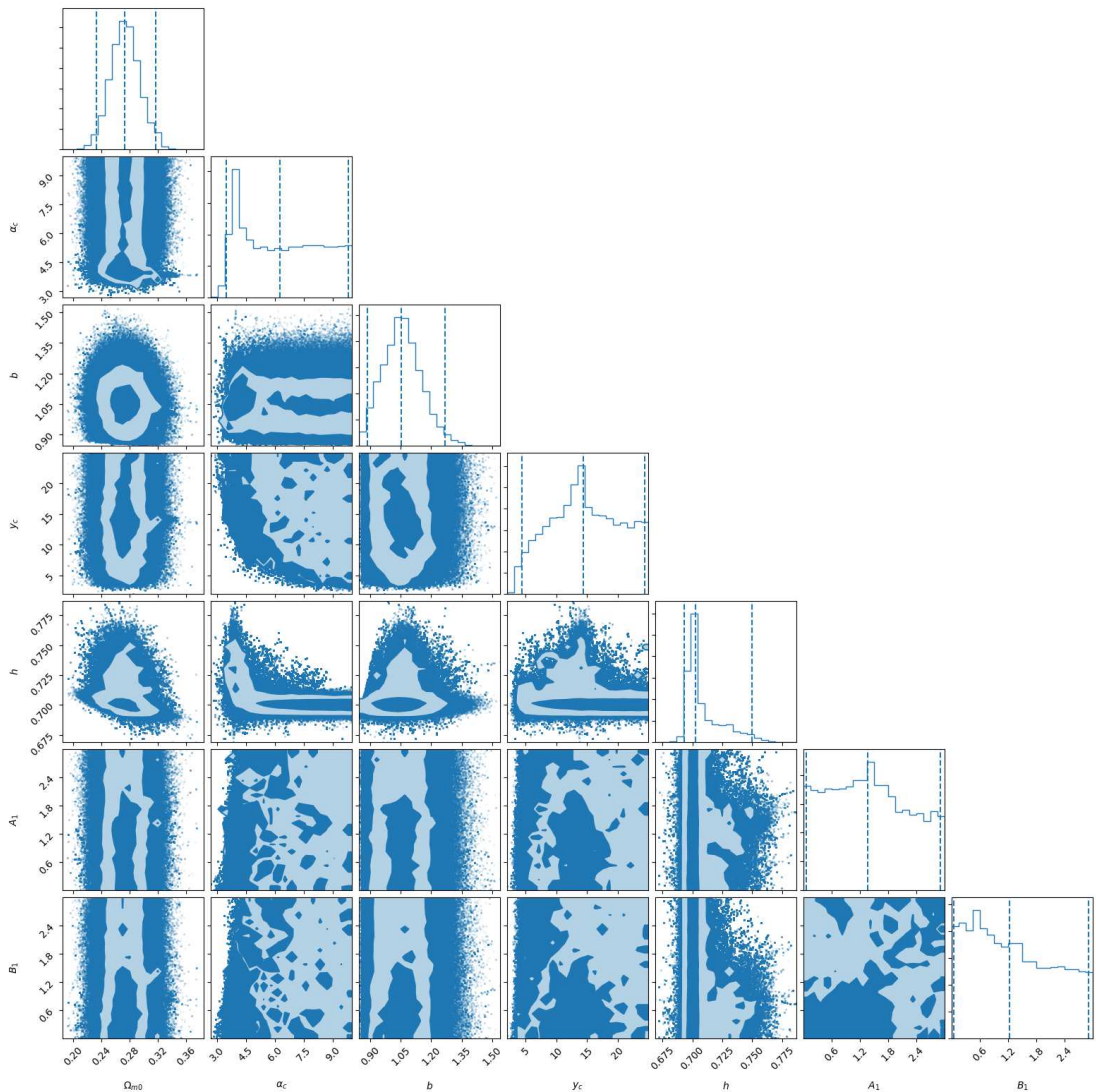}
\ec
\caption{\label{fig1} Marginalized constraints on the parameters $\Om_{{\rm m}0}$, $\a_{\rm c}$, $b$, $y_{\rm c}$, $h$, $A_1$ and $B_1$ for the theory with weighted derivatives with multiharmonic modes, $N=2$ and $A_0=0$. Although there is degeneracy in some of the parameters, this model is able to reach higher values of $h$.}
\end{figure*}

\begin{figure}[ht]
\bc
\includegraphics[width=0.95\columnwidth]{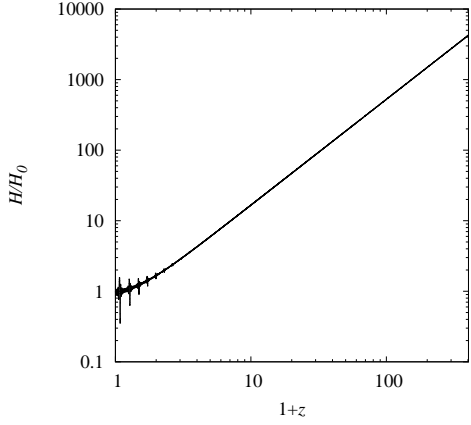}
\ec
\caption{\label{fig2} Behavior of $H/H_0$ for the multiharmonic best fit with $N=2$ and $A_0=0$. Oscillations start at low redshift values. The plot for $A_0=1$ is the same.}
\end{figure}

\begin{figure}[ht]
\bc
\includegraphics[width=0.95\columnwidth]{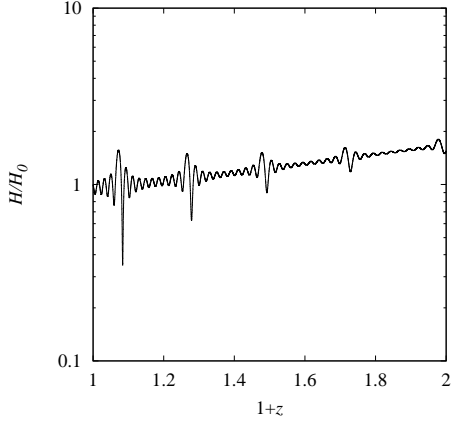}
\ec
\caption{\label{fig3} Behavior of $H/H_0$ for the multiharmonic best fit with $N=2$ and $A_0=0$ at low redshifts. The plot for $A_0=1$ is the same.}
\end{figure}

\begin{figure}[ht]
\bc
\includegraphics[width=0.9\columnwidth]{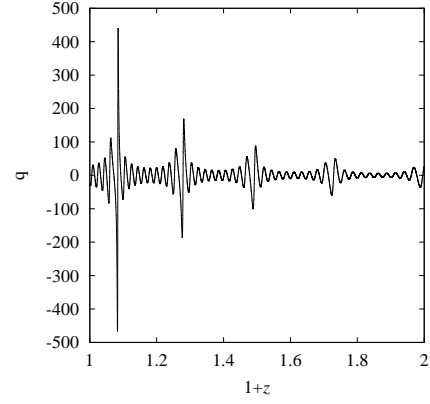}
\ec
\caption{\label{fig4} Behavior of the deceleration parameter q for the multiharmonic best fit with $N=2$ and $A_0=0$. We can see the ``heartbeat'' of dark energy playing a nontrivial role at late times. The plot for $A_0=1$ is the same.}
\end{figure}

\begin{figure}[ht]
\bc
\includegraphics[width=0.9\columnwidth]{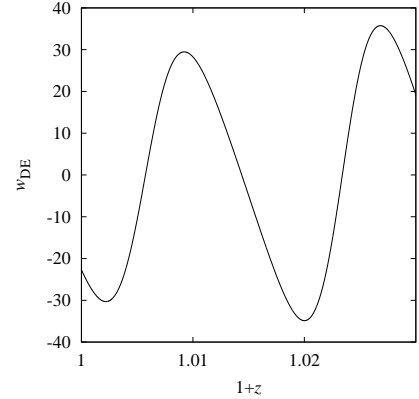}
\ec
\caption{\label{fig5} Behavior of $w_\textsc{de}$ for the multiharmonic best fit with $N=2$ and $A_0=0$. The plot for $A_0=1$ is the same. Thanks to the last oscillation, the universe is able to increase $H$ to values close to the measured value $H_0$ today. Here a dark-energy epoch is reached due to oscillations which make the universe accelerate and decelerate alternatively. Acceleration becomes then a transient but today the universe does undergo a period of accelerated expansion.}
\end{figure}

\begin{figure*}[t]
\bc
\includegraphics[width=15cm]{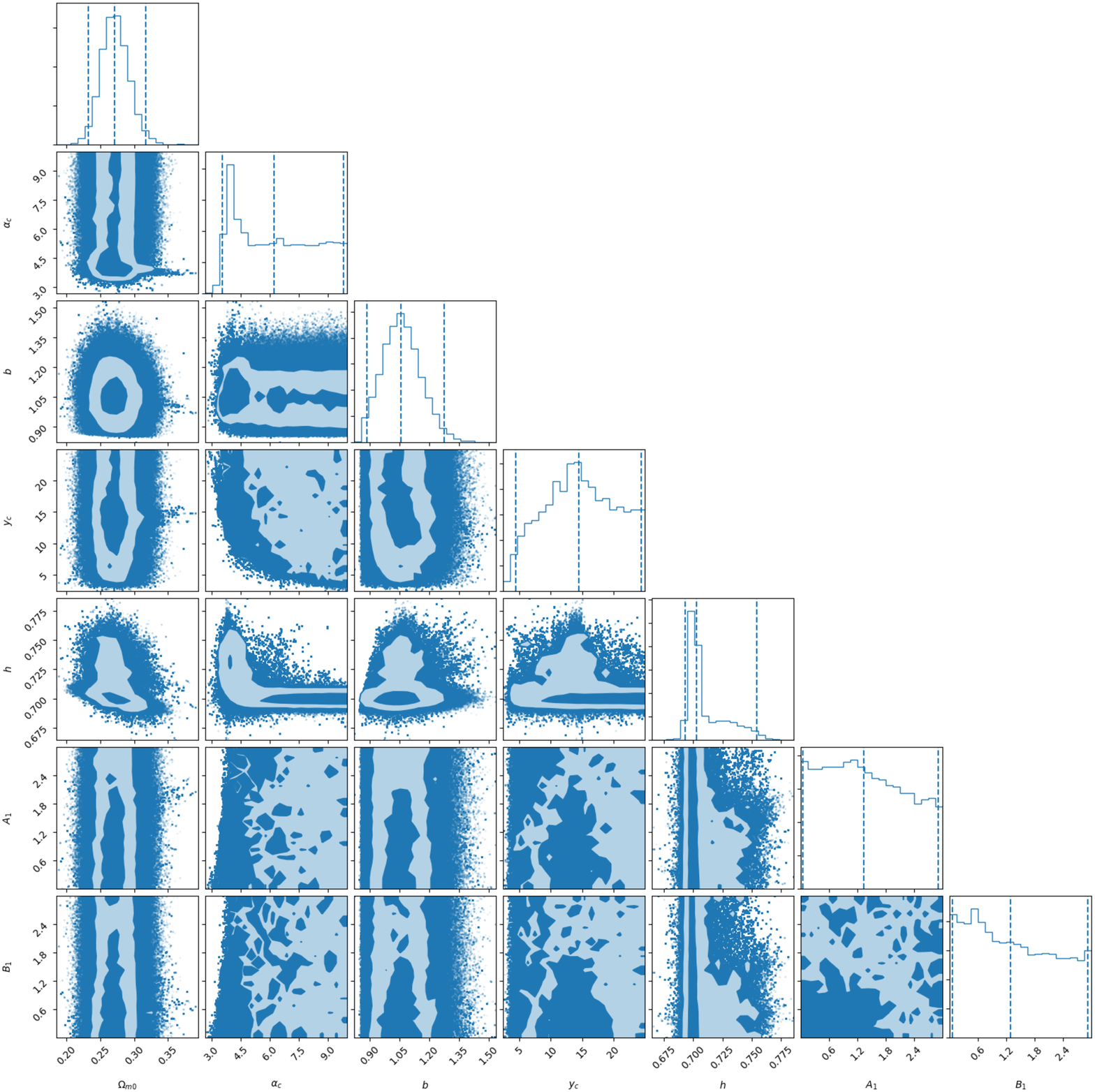}
\ec
\caption{\label{fig6} Marginalized constraints on the parameters $\Om_{{\rm m}0}$, $\a_{\rm c}$, $b$, $y_{\rm c}$, $h$, $A_1$ and $B_1$ for the theory with weighted derivatives with multiharmonic modes, $N=2$ and $A_0=1$.}
\end{figure*}

\begin{itemize}
\item \textbf{Case 1: no oscillations}. In the absence of log oscillations, $A_0=1$ and the model has only five free parameters: $\a_{\rm c}$, $y_{\rm c}$, $b$, $h:=H_0/(100\,{\rm km}\,{\rm s}^{-1}\,{\rm Mpc}^{-1})$ and $\Om_{{\rm m}0}$. As priors, we used $-3\leq \alpha_{\rm c} \leq 10$, $10^{-5}\leq y_{\rm c} \leq 20$, $0.1\leq b\leq 3$, $0.6\leq h\leq 0.8$, $0.1\leq \Omega_{\rm m 0}<0.5$. The marginalized likelihood 2$\sigma$ contours in the parameter space ($2\s$ preferred values of the parameters) are shown in Table \ref{tab1}.
 We will discuss the physical implications of the results of this and the other cases in Sec.\ \ref{disc}. For the time being, note that the value of $h$ is in tension with the local observations of $H_0$.
\item \textbf{Case 2a: one harmonic, $\bm{N=2}$}. The above results do not change in the presence of only one harmonic mode. As priors, we used the same as above plus $0<A_1,B_1<1$. The results for $A_0=0$ and $A_0=1$ are in Table \ref{tab1}.
 For some data, there is a large degeneracy, especially in the parameters of the $v$ function, namely $y_{\rm c}$, $\alpha_{\rm c}$, $A_1$ and $B_1$. On the other hand, the value of $h$ is strongly constrained and it is still in tension with the local observations of $H_0$. This situation seems not to depend on the value of $N$. When $A_0=1$, the upper bound on $\Omega_{\rm m 0}$ slightly increases.
\item \textbf{Case 2b: one harmonic, $\bm{N=\rme^{10}}$}. Having chosen the same priors as before, we find the results of Table \ref{tab1}.
 Again, the value of $h$ is in tension with today's value of the Hubble parameter and the upper bound of $\Omega_{\rm m 0}$ increases when $A_0=1$.
\item \textbf{Case 3a: many harmonics, $\bm{N=2}$}. This is the most promising case.
 The likelihood contours for $A_0=0$ are shown in Fig.~\ref{fig1}. Notice in Table \ref{tab1} the higher value of $h$ and the lower bound at 2$\s$ for $\alpha_{\rm c}$. In order to explore this better fit more in detail, we tried to enlarge the prior for $A_1$ and $B_1$ to $0<A_1,B_1<3$. However, this resulted in a large degeneracy for these two parameters. The $\chi^2$ for this case ($\chi^2_{\rm min}=555$) is significantly smaller than the one for all the previous cases (values approximately equal to 565). The behavior of the Hubble factor is shown in Fig.\ \ref{fig2}, while a more detailed dynamics for $H/H_0$ at low redshifts is depicted in Fig.\ \ref{fig3}. At the same time, it is interesting to show also the deceleration parameter ${\rm q}$ (Fig.\ \ref{fig4}) and $w_\textsc{de}$ (Fig.\ \ref{fig5}). Thanks to the oscillations, we are able to get a higher value for $H_0$, although acceleration might be transient, i.e., existing today but stopping some time in the future. The results for $A_0=1$ are given in Fig.~\ref{fig6}. The dynamics of this case in terms of the Hubble factor, $H/H_0$ at low redshifts, q and $w_\textsc{de}$ is virtually indistinguishable from that shown in Figs.\ \ref{fig2}--\ref{fig5}.
\item \textbf{Case 3b: many harmonics, $\bm{N=\rme^{10}}$}. Here one can appreciate a difference between the $N=2$ and the $N=\rme^{10}$ results, the latter getting considerably worse. Once more, the $\chi^2$ for the minimum shifts to values close to 565. 
\end{itemize}
\begin{table*}[t]
\centering
\renewcommand{\arraystretch}{1.2}
\setlength\arrayrulewidth{0.5pt}
\begin{tabular}{lccc|ccccc}\hline\hline
Case    & $n_{\rm max}$ & $N$ & $A_0$ & $\a_{\rm c}$ ($\a_{\rm c}^{\rm min}$) & $y_{\rm c}^{\rm min}$ & $h$ & $b$ & $\Om_{{\rm m}0}$ \\\hline
{\bf 1} & $0$	& --- & $1$ & $5.1$ ($0.6$) & $1.7$ & $0.70^{+0.01}_{-0.01}$ & $1.04^{+0.21}_{-0.17}$ & $0.278^{+0.200}_{-0.400}$ \\\hline
\multirow{2}{*}{{\bf 2a}} & \multirow{2}{*}{$1$} & \multirow{2}{*}{$2$} & $0$ & $7.0$ ($3.4$) & $3.2$ & $0.70^{+0.01}_{-0.01}$ & $1.05^{+0.20}_{-0.17}$ & $0.270^{+0.040}_{-0.040}$ \\
& & & $1$ & $6.7$ ($0.3$) & $2.0$ & $0.70^{+0.01}_{-0.01}$ & $1.04^{+0.22}_{-0.16}$ & $0.275^{+0.200}_{-0.040}$ \\
\multirow{2}{*}{{\bf 2b}} & \multirow{2}{*}{$1$} & \multirow{2}{*}{$\rme^{10}$} & 0 & $5.7$ ($0.4$) & $1.7$ & $0.70^{+0.01}_{-0.01}$ & $1.05^{+0.20}_{-0.17}$ & $0.270^{+0.050}_{-0.080}$ \\
& & & $1$ & $5.6$ ($1.0$) & $1.9$ & $0.70^{+0.01}_{-0.01}$ & $1.04^{+0.21}_{-0.17}$ & $0.276^{+0.130}_{-0.050}$ \\
\rowcolor{gray!16} &  &  & $0$ & $3.8$ ($3.7$) & $3.9$ & $0.73^{+0.01}_{-0.03}$ & $1.05^{+0.21}_{-0.17}$ & $0.276^{+0.036}_{-0.040}$ \\
\rowcolor{gray!16}\multirow{-2}{*}{{\bf 3a}} & \multirow{-2}{*}{$10$}& \multirow{-2}{*}{$2$} & $1$ & $3.8$ ($3.6$) & $4.1$ & $0.73^{+0.02}_{-0.04}$ & $1.05^{+0.22}_{-0.17}$ & $0.276^{+0.040}_{-0.044}$ \\
\multirow{2}{*}{{\bf 3b}} & \multirow{2}{*}{$10$} & \multirow{2}{*}{$\rme^{10}$} & 0 & $7.2$ ($3.7$) & $4.0$ & $0.70^{+0.01}_{-0.01}$ & $1.05^{+0.22}_{-0.17}$ & $0.273^{+0.04}_{-0.04}$ \\
& & & $1$ & $7.1$ ($3.4$) & $4.0$ & $0.70^{+0.01}_{-0.01}$ & $1.05^{+0.21}_{-0.17}$ & $0.273^{+0.046}_{-0.038}$ \\\hline\hline
\end{tabular}
\caption{\label{tab1} Preferred values of the parameters $h$, $b$ and $\Om_{{\rm m}0}$ with $2\s$ (95\,\% confidence level) errors. For $\a_{\rm c}$ we indicate the preferred value and the $2\s$ lower bound $\a_{\rm c}^{\rm min}$, while for $y_{\rm c}$ we only show the lower bound $y_{\rm c}^{\rm min}$. The values of $A_1$ and $B_1$ are not shown due to strong degeneracy. The shaded row corresponds to the only case reproducing the observed late-time Hubble parameter.}
\end{table*}



\subsection{Consequence: Extending multiscale spacetimes}\label{disc}

All of the above cases share some common features. The value of the parameter $b$ is close to 1, which means that the inverse Hubble parameter today is approximately equal to the age of the universe, just like in the standard $\Lambda$CDM model: $t_0 H_0\approx 1$. Similarly, $\Om_{{\rm m}0}$ is close to the $\Lambda$CDM value.

The value of $h$ is the most important discriminator selecting viable models. Only those with many harmonics and $N=2$ (case 3a) are able to recover the estimated value from late-time observations.\footnote{Since we have not used early-universe data, we are not in a position to say anything about the $H_0$ tension.} Concentrating only on case 3a, the lower bound on $y_{\rm c}$ is
\be\label{yst}
y_{\rm c}=\frac{t_{\rm c}}{t_0}>3.9\,,
\ee
which means that the characteristic time $t_{\rm c}$ is always much larger than the age of the universe $t_0$. On one hand, this is compatible with the finding that the value of $A_0$ has little impact on the numerical analysis, i.e., that the zero mode is strongly suppressed and hence it does not affect cosmology. On the other hand, the bound \Eq{yst} is not in contradiction with the fact that multifractional effects can explain observations at times $O(t_0)$, since the modulation of superposing log oscillations of the highest frequency $\om$ takes place at much shorter scales. Note the specifications of ``superposing'' and ``highest frequency'': in the absence of log oscillations (case 1), in the presence of only one harmonic ($n_{\rm max}=1$, cases 2a and 2b), or when the harmonics frequency is too low (large $N$, cases 2b and 3b), the theory is unable to recover the observed late-time value of the Hubble parameter. Note the importance of having a theoretical upper limit on $\om$. Without it, it would have been more difficult to find a modulation comparable with the observable patch of the universe and compatible with late-time data.

Concerning the value of the oscillation amplitudes $A_1$ and $B_1$,  as one can see in Figs.~\ref{fig1} and \ref{fig6} their distribution is uniform inside the prior $0<A_1,B_1<3$. Therefore, there is a strong degeneracy in the parameter space and it is not possible to establish a preferred value with the late-time data we used.

Our results indicate that the theory with weighted derivatives can serve not only as a microscopic (UV) modification of gravity, but also as a cosmological (IR) one. However, unlike other IR modifications of standard cosmology, this theory does not rely on the typical UV/IR divide through a characteristic scale. When the multifractional geometry is of this type (no log oscillations), it fails to fit data, since the scale $t_{\rm c}$ is larger than the age of the universe. Thus, we do \emph{not} end up with the typical model where general relativity is an adequate description of the cosmos during most of the history of the universe and the recent acceleration is triggered around a certain IR scale.

Rather, multifractional effects embodied in logarithmic oscillations are endemic to the cosmological picture of the theory and mimic, in a subtle way, a standard cosmology with a late-time dominating dark-energy component. Here the characteristic scale is very large (actually, beyond the Hubble horizon) but, nevertheless, it modifies the evolution of the universe via the harmonic structure of the geometry.

The last piece of information gathered about the geometry of the cosmos reserves a surprise and is about the Hausdorff dimension of spacetime. Just as in the case of $y_{\rm c}$, the distribution of values of $\a_{\rm c}$ is flat past the peak (preferred value) and one cannot establish an upper bound. However, while the peak of $y_{\rm c}$ is rather mild, the one for $\a_{\rm c}$ is very pronounced and allows us to clearly establish a preferred value. In case 3a, the latter is
\be\label{a38}
\a_{\rm c}\approx 3.8\,,
\ee
with a very small error bar. The preferred value of $\a_{\rm c}$ is larger in the other cases. This means that the theory with weighted derivatives cannot accommodate at the same time particle-physics constraints and explain the late-time acceleration of the universe if we consider only the binomial measure \Eq{bino} [for which $t_*=t_{\rm c}$ and \Eq{yst} would be in gross contradiction with \Eq{tstt0}], while the case with many harmonics of maximal frequency can if we allow for an ultra-IR term as in \Eq{bino2}. 

The result \Eq{a38} has a clear geometric interpretation based on the calculation of the Hausdorff dimension in a perfectly homogeneous FLRW background. On this spacetime, the measure weight is \Eq{genpo} but, by definition \Eq{dhdef}, the time Hausdorff dimension $\dh^{\rm time}$ is computed after averaging over log oscillations \cite{revmu,frc2}, i.e., we must take \Eq{bino2} instead of its version with log oscillations. When $A_0=0$, the averaged measure is $v=1$ and $\dh^{\rm time}=1$ at all times. However, when $A_0=1$ the Hausdorff dimension of time is no longer constant. With a FLRW metric and $A_0=1$, the effective action is one-dimensional,
\be
S_{\rm FLRW}=\int\rmd t\,v(t)\,a^3(t)\,\cL
\ee
and the volume from \Eq{bino2} (omitting the metric density term, i.e., curvature effects\footnote{These are always excluded when computing the Hausdorff or spectral dimension.}) is
\be
\cV(t)=\int^t\rmd t'\,v(t')=\frac{1}{\a_*}\left|\frac{t}{t_*}\right|^{\a_*}+t+\frac{1}{\a_{\rm c}}\left|\frac{t}{t_{\rm c}}\right|^{\a_{\rm c}}\,,
\ee
so that
\be\label{dh2}
\dh^{\rm time}=\frac{|t/t_*|^{\a_*}+t+|t/t_{\rm c}|^{\a_{\rm c}}}{(1/\a_*)|t/t_*|^{\a_*}+t+(1/\a_{\rm c})|t/t_{\rm c}|^{\a_{\rm c}}}\,.
\ee
Assuming $0<\a_*<1$ and $\a_{\rm c}>1$, when $t\ll t_*$ one has $\dh^{\rm time}\simeq \a_*$. At intermediate scales $t_*\ll t\lesssim t_{\rm c}$, the Hausdorff dimension of time coincides with the topological dimension, $\dh^{\rm time}\simeq 1$. This is the regime characterizing most of the history of the universe. In a late-time regime $t\gg t_{\rm c}$ posterior to the present epoch, one has $\dh^{\rm time}\simeq\a_{\rm c}$.

Our finding is that $\a_{\rm c}\approx 4$ in order to reproduce the observed late-time acceleration. Here, dark energy is not an extra matter component in the universe: it is the manifestation of a dimensional flow where the limiting value of the time Hausdorff dimension is the topological dimension 4, as if dimensional flow were recovering the spatial dimensions ``lost'' in the symmetry reduction from a generic background to a perfectly homogeneous spacetime. Of course, the FLRW dynamics is not really blind to the value of the topological dimension: the value $\a_{\rm c}=4$ is implicitly induced by the $D=4$ factors hidden in the Friedmann and continuity equations. However, the source of this numerical coincidence is not obvious.

We hereby propose the following picture. At times of order of the Planck scale, the universe is dominated by quantum-gravity effects of UV  type. In general, quantum gravity is associated with dimensional flow, which, in turn, is described by a generalized polynomial measure weight \Eq{genpo} \cite{revmu,first}. In the literature, only UV terms have been considered in this polynomial expansion, but quite generally also IR terms can be conceived. Therefore, the time part $v(t)$ of the spacetime measure weight $v(x)=v(t)\,v_1(x^1)\,v_2(x^2)\,v_3(x^3)$ would be \Eq{bino2} instead of \Eq{bino}. Similar expressions hold in the spatial directions with $t$, $t_*$, $\a_*$, $t_{\rm c}$ and $\a_{\rm c}$ replaced by $x^i$, a spatial scale $\ell_*$, a set of fractional exponents $\a_i$, and so on. 

To study the dimension of spacetime \Eq{dhdef} we have to average out the log oscillations according to the definition \Eq{avef}, which is valid only in the presence of one frequency. In the case of the trinomial measure, we have two sets of frequencies governed by the fractional exponents in \Eq{omst} and \Eq{omspe}. According to the regime considered, one will use one or the other frequency.

At very early times $t\lesssim t_*$, $v(t)\simeq |t_*/t|^{1-\a_*}$, short-scale quantum-gravity effects dominate and the Hausdorff dimension of spacetime is given by \Eq{dh}. This is the usual regime where multifractional theories have been studied \cite{revmu,frc16}. Here we are in an early, pre-inflationary regime where our causal patch of the universe is inhomogeneous and anisotropic. These effects quickly die away before or during inflation: the residual IR correction in \Eq{bino2} survives but is negligible at times $t_*<t\ll t_{\rm c}$ and the universe is described by general relativity to a good approximation, $v(t)\simeq 1$. If this transition occurs before inflation or in inhomogeneous regions, the local Hausdorff dimension of spacetime is $\dh^{\rm spacetime}=4$. The onset of inflation makes the causal patch approximately homogeneous and isotropic until our times, where, however, IR multiscale effects become more and more important and $v(t)\simeq |t/t_{\rm c}|^{\a_{\rm c}-1}$. The evolution of the effective Hausdorff dimension of spacetime in these three snapshots of the universe (quantum-gravity dominated, general-relativistic and late-time accelerating) is summarized in Table \ref{tab2}.
\begin{table}[t]
\begin{center}
\begin{tabular}{lccc}\hline\hline
Regime             & $v(t)$ 													& $\dh^{\rm time}$ & $\dh^{\rm spacetime}$   \\\hline
Quantum gravity    & $|t_*/t|^{1-\a_*}$ 							& $\a_*$   				 & $\a_*+\sum_{i=1}^3\a_i$ \\
General relativity & 1						& 1								 & $1+3=4$								 \\ 
\quad (general background) & & & \\
General relativity & 1													& 1								 & 1											 \\ 
\quad (FLRW) & & & \\
Late times (FLRW)	 & $|t/t_{\rm c}|^{\a_{\rm c}-1}$		& $\a_{\rm c}=4$ 	 & $\a_{\rm c}=4$ 				 \\\hline\hline
\end{tabular}
\end{center}
\caption{\label{tab2} Time and spacetime Hausdorff dimension in different regimes, after averaging out log oscillations.}
\end{table}

\begin{figure}[t]
\bc
\includegraphics[width=0.8\columnwidth]{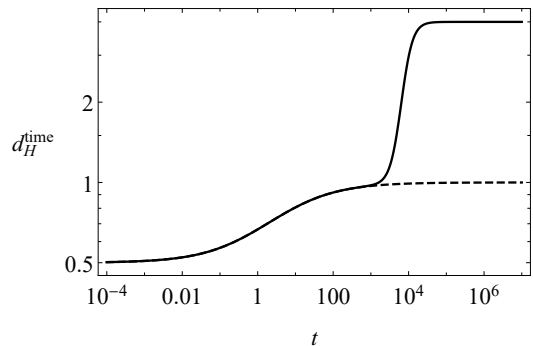}
\ec
\caption{\label{fig7} Hausdorff dimension of time \Eq{dh2} (solid curve) in the theory with weighted derivatives, for $t_*=1$, $\a_*=1/2$, $t_{\rm c}=600$ and $\a_{\rm c}=4$. The Hausdorff dimension of time $\dh^{\rm time}=(|t/t_*|^{\a_*}+t)/[(1/\a_*)|t/t_*|^{\a_*}+t]$ in the absence of the ultra-IR term (dashed curve) is shown for comparison.}
\end{figure}

Notice that the third phase in the table (general relativity on a FLRW background) is not necessary. In fact, the extended multifractional model with weighted derivatives accounts for the whole history of the universe from post-inflation until today, without adding any dark-energy component by hand. However, if $t_{\rm c}\gg t_*$ the intermediate general-relativity phase will show up anyway during the history of the universe. An example is given by the profile \Eq{dh2} of the Hausdorff dimension of time shown in Fig.\ \ref{fig7}. The fourth and last phase (late-time multifractional FLRW evolution) restores the value $\dh^{\rm spacetime}=4$ of the Hausdorff dimension of spacetime. Thus, even if $\dh^{\rm spacetime}$ is not conserved during the evolution of the universe (there would be no dimensional flow in that case), it is restored asymptotically if the universe stays exactly homogeneous.

Note that the regimes listed in Table \ref{tab2} are phenomenologically mismatched with respect to the observability window of multifractional effects. That is, since $t_{\rm c}\gg t_0$, we cannot observe time dimensional flow now because we are in the general-relativity plateau, far away both from the UV and the ultra-IR regimes of the profile $\dh(t)$. However, the inception of cosmic acceleration happens well before the establishment of the ultra-IR regime. We are not aware of similar mechanisms in any other models of dark energy.


\section{Conclusions}\label{conc}

Multifractional spacetimes have been previously explored in a range of settings and regimes, from particle physics to cosmology. Here we applied them to the late-time universe in order to see what they can say about the dark-energy problem and, conversely, how supernov\ae\ data can constrain such models. While the theory with $q$-derivatives cannot explain late-time acceleration, the theory with weighted derivatives has a chance. To do so, we considered a term in the spacetime measure that describes a sort of ultra-IR regime beyond Einstein gravity. This term was not added \emph{ad hoc} in the measure expansion: it was already there, admitted by the dimensional-flow theorem \cite{first}, but in the literature it had been overlooked because this theory was studied mainly in a microscopic particle physics setting (e.g., \cite{frc16}) where such term is completely negligible. In this sense, we are not proposing a ``new'' class of models but simply continuing the study of old ones in previously untapped regimes.

The result is that none of the simplified versions of the multifractional theory with weighted derivatives explored more frequently in the literature, the one with monotonic measure (no log oscillations) and the one with only one harmonic, can explain late-time data. Only when enough undamped harmonics are present can the theory explain the late-time acceleration of the universe without fine tuning and without invoking extra dynamical degrees of freedom. For both $\alpha_{\rm c}$ and $y_{\rm c}$ we only have an $O(1)$ lower bound for all the models, that is, a large degeneracy on those two variables. A fine tuning in this context would correspond either to lower bounds $\alpha_{\rm c},y_{\rm c}\gg 1$ or to an allowed region with negligible area in the two-dimensional parameter space. In contrast, to explain dark energy with a cosmological constant in general relativity a fine tuning of tens of digits on the value of $\Lambda$ is required.

If the oscillation amplitudes are sufficiently large, the effective dark energy density log periodically dominates over the matter energy density, while the effective barotropic index $w_\textsc{de}$ log periodically oscillates between positive and negative values  infinitely many times during the history of the universe. The last oscillation causes a late-time acceleration period (Fig.\ \ref{fig5}), but not as an isolated event: it is the last but the most pronounced of an infinite sequence of acceleration epochs (Fig.\ \ref{fig4}).

In turn, data suggest that the preferred value of the Hausdorff dimension of spacetime in the ultra-IR regime is tantalizingly close to 4, which stimulates further theoretical research on the spacetime geometry of this theory along the lines discussed in Sec.\ \ref{disc}. It will be worth investigating further the geometric interpretation presented therein, due to its important phenomenological consequences as an alternative explanation of late-time acceleration. The absence of any fine tuning in the parameters of the theory could make it an appealing alternative to other geometry-driven or matter-field-driven explanations of dark energy. Future experiments in addition to the search of other constraints for these model will shed further light into the restoration law proposed here.


\begin{acknowledgments}
A.D.F.\ was supported by JSPS KAKENHI Grant No.\ 20K03969. G.C.\ is supported by the I+D grant FIS2017-86497-C2-2-P of the Spanish Ministry of Science, Innovation and Universities and acknowledges the contribution of the COST Action CA18108. He also thanks the Yukawa Institute for kind hospitality during part of the development of this project.
\end{acknowledgments}


\end{document}